\documentclass[aps,amssymb,amsfonts,amsmath,preprint]{revtex4}
\usepackage{hyperref}

\usepackage{graphicx}
\usepackage{dcolumn}
\usepackage{bm}

\begin{document}

\title{Plasmon spectra of nanospheres under a tightly focused beam}

\author{Nassiredin M. Mojarad}
\author{Vahid Sandoghdar}
\author{Mario Agio}
\email{mario.agio@phys.chem.ethz.ch}
\affiliation{Nano-Optics Group, Laboratory of Physical Chemistry, ETH Zurich
\\  8093 Zurich, Switzerland}

\begin{abstract}
We study the modification of the far-field cross sections and the 
near-field enhancement for gold and silver nanospheres 
illuminated by a tightly focused beam. Using a multipole-expansion 
approach we obtain an analytical solution to the scattering problem and
provide insight on the effects of focusing on the optical response.
Large differences with respect to Mie theory are found especially when the
nanoparticle supports quadrupole or higher-order resonances.
\end{abstract}


\maketitle 

\section{INTRODUCTION}
The optical properties of metal nanoparticles 
(NPs)~\cite{bohren83,kreibig95} 
represent a topic of active research in several areas, like 
nano-optics~\cite{kalkbrenner05,kuehn06,anger06}, 
solid-state physics~\cite{voisin00,arbouet03,stoller06}, 
biology~\cite{schultz00,fritzsche03,seelig07} and 
photonics~\cite{maier03,waele07}. What makes them particularly attractive
is their ability to sustain an electromagnetic resonance yet being
much smaller than the incident wavelength. This peculiar feature 
originates from the existence of an electron-density mode that can 
couple to external radiation giving rise to a localized 
surface-plasmon-polariton resonance (SPR)~\cite{bohren83,kreibig95}.
In the experimental practice, metal NPs have been often studied 
close to interfaces and under both wide-field and tight 
illumination~\cite{soennichsen00,mock03,linfords04,berciaud04}. 
Nevertheless, since the NP is very small compared to the wavelength, the 
measured plasmon spectrum is usually only compared to theory
based on plane wave (PW) illumination~\cite{bohren83}. 
While the effect of supporting films and substrates has been extensively 
investigated~\cite{quinten99,arias01,videen05,aslan05,moreno06},
to the best of our knowledge, there is no detailed study on the spectrum 
of metal NPs under tight illuminations~\cite{torok98,challener03}.

The electromagnetic problem of a linearly-polarized PW
scattered by a spherical particle dates back to
the work of Mie~\cite{mie08}. After the invention of the laser,
the availability of narrow band and collimated beams has posed new 
theoretical issues on the scattering of light by particles.
Morita et al.~\cite{morita68}, and Tsai and Pogorzelski~\cite{tsai75}
were the first to study the case of spherical particles illuminated by
a Gaussian beam whose waist is smaller or larger than the
particle size. A decade later, Barton et al.~\cite{barton88}
extended the study from the far field to the near and internal 
fields. Other authors have also considered the scattering properties of 
objects under Gaussian illumination~\cite{tam78,gouesbet88,lock95,gouesbet99}.
In summary, all works have found that there is
a significant deviation from the case of PW illumination
if the particle size is larger than the beam waist.

Less attention has been given to the case of a high-numerical-aperture 
(high-NA) beam incident upon a metal NP~\cite{torok98,challener03}.
T\"or\"ok et al.~\cite{torok98} and Challener et 
al.~\cite{challener03} have described the high-NA beam by PW 
expansion~\cite{richards59}. 
Subsequently, the scattering problem was solved using classical Mie 
theory~\cite{torok98} or the finite-difference time-domain 
method~\cite{challener03}.
Although the PW expansion provides an accurate description of
the electromagnetic field at the focus, it does not provide an easily 
accessible physical insight on the effect of strong focusing. 
Moreover, in the mentioned 
studies, only selected wavelengths were considered and the modification 
of the plasmon spectrum was not taken into account.

The aim of this work is to study the interaction of a high-NA beam
with metal NPs in more detail and compare it with the case of PW
illumination. By solving the scattering problem using a multipole-expansion
approach~\cite{sheppard97}, we gain intuition on the NP's 
optical response  both in the far and near field. 
We show that even when the NP is smaller than the focal spot,
the difference in the multipole content of a high-NA beam compared to 
a PW causes a modification of the plasmon spectrum, especially
when the metal NP exhibits resonances of order higher than the dipolar one.

The paper is organized as follows. In Sec.~\ref{theory} we present the 
analytical solution of generalized Mie theory~\cite{gouesbet82} for a 
high-NA beam incident on a spherical NP placed at the focus.
We also derive formulae for the scattering and extinction cross 
sections and for the average intensity enhancement in the near field. In  
Sec.~\ref{results} we present and discuss results for gold and silver 
NPs with a 100~nm diameter.

\section{THEORY}
\label{theory}
\subsection{Generalized Mie Theory for a High-NA Beam}

\begin{figure}[h]
\begin{center}
\includegraphics[width=8.3cm]{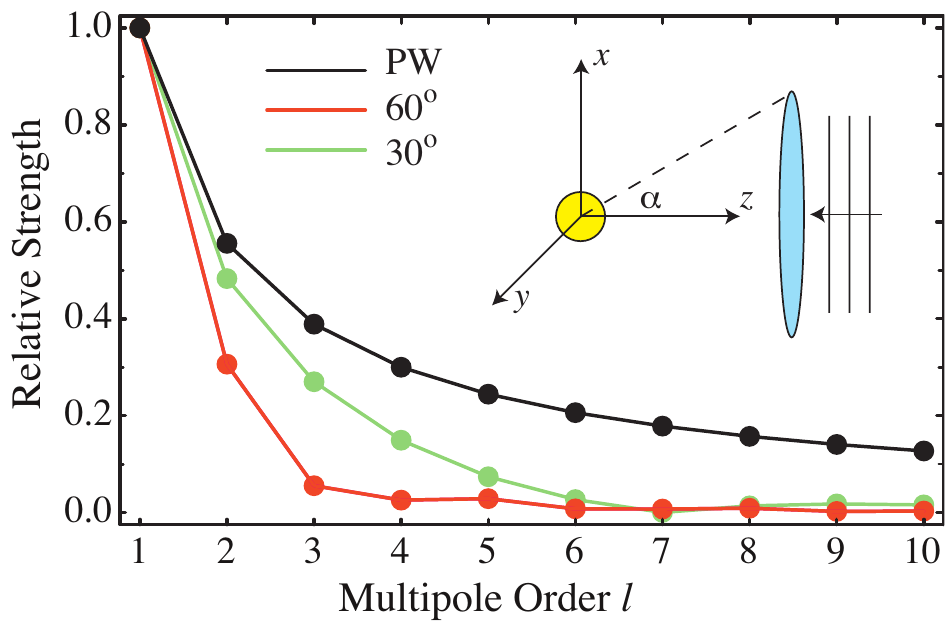}
\caption{\label{multipole}
Relative strength of the multipole coefficients for the expansion 
of a PW $|E_l/E_1|$ and a high-NA beam 
($\alpha=60^\mathrm{o}$ and $\alpha=30^\mathrm{o}$) 
$|A_l/A_1|$. Inset: a PW polarized along $x$ and propagating along $-z$
is focused by a high-NA lens and scatters on a metal NP
positioned at the focus. $\alpha$ is the angular semi-aperture
of the lens.}
\end{center}
\end{figure}

The generalized Mie theory of a high-NA beam incident on a spherical
NP placed at the focus is carried out by 
expanding the electromagnetic field in multipoles.
The inset to Fig.~\ref{multipole} sketches the layout of the problem.
A PW polarized along $x$ and propagating along $-z$ is focused
by a high-NA aplanatic system~\cite{richards59}.
It can be shown that the incident field components at the reference 
sphere in the image space are~\cite{richards59,sheppard97}
\begin{eqnarray}
\label{esph1}
E_{\mathrm{i},r} & = & 0,\\
\label{esph2}
E_{\mathrm{i},\theta} & = &  E_0e^{-ikf}a(\theta)\cos\phi,\\
\label{esph3}
E_{\mathrm{i},\phi  } & = & -E_0e^{-ikf}a(\theta)\sin\phi,
\end{eqnarray}
in the spherical coordinates ($r$, $\theta$, $\phi$).
Here $E_0$ is the PW amplitude and $k$ is the wavevector.
The parameter $a(\theta)$ represents the illumination angular weighting 
factor and $f$ is the lens focal length. The factor $\exp(-ikf)$
is added to ensure phase agreement with the electric field computed by 
PW expansion~\cite{richards59}.
For a system satisfying the sine condition one finds 
$a(\theta)=(\cos\theta)^{1/2}$~\cite{richards59}.

The multipole-expansion coefficients of the incident field in the image space
are obtained by enforcing the boundary conditions set by
Eqs.~(\ref{esph1}-\ref{esph3})~\cite{sheppard97}.
The $\phi$ dependence in these equations 
already tells us which multipoles can properly represent 
the incident field, namely
\begin{equation}
\label{eincexp}
\mathbf{E}_\mathrm{i}(r,\theta,\phi)=\sum_l\left[A_l
\mathbf{N}_{\mathrm{e}1l}(r,\theta,\phi)+B_l
\mathbf{M}_{\mathrm{o}1l}(r,\theta,\phi)\right],
\end{equation}
where the sum over $l$ extends from 1 to infinity and
$\mathbf{N}_{\mathrm{e}1l}$, $\mathbf{M}_{\mathrm{o}1l}$ 
are the vector spherical harmonics defined in Ref.~\onlinecite{bohren83}.
The explicit dependence of the electromagnetic field on ($r,\theta,\phi$) 
is omitted in the following expressions for the sake of brevity.
Furthermore, we choose the spherical Hankel function of the second kind
($h^{(2)}_l(\rho)$)~\cite{jackson99} in the multipoles
so that the incident field is an incoming wave. Using the asymptotic 
expression $h^{(2)}_l(\rho)\simeq i\exp[-i(\rho-\pi l/2)]/\rho$ and 
matching the far fields of Eq.~(\ref{eincexp}) to 
Eqs.~(\ref{esph1}-\ref{esph3}), we obtain the conditions $B_l=-iA_l$ and
\begin{equation}
\sum_l i^lA_l\left[\frac{P_l^1(\cos\theta)}{\sin\theta}+
\frac{\mathrm{d}P_l^1(\cos\theta)}{\mathrm{d}\theta}\right]=
kfE_0a(\theta),
\end{equation}
where $P_l^1(\cos\theta)$ are associated Legendre functions~\cite{jackson99}
and $\rho=kf$.
Using the procedure described in Ref.~\onlinecite{sheppard97}, the
expansion coefficients are found to be
\begin{equation}
A_l=(-i)^lE_0kf\frac{2l+1}{2l^2(l+1)^2}\int_0^\alpha a(\theta)
\left[\frac{P_l^1(\cos\theta)}{\sin\theta}+
\frac{\mathrm{d}P_l^1(\cos\theta)}{\mathrm{d}\theta}\right]
\sin\theta\mathrm{d}\theta.
\end{equation}
The parameter $\alpha$ is the angular semi-aperture of the lens and
is determined by the formula NA=$n_\mathrm{b}\sin\alpha$,
where $n_\mathrm{b}$ is the refractive index of the background medium.
The regularity of the electromagnetic field at the
focus requires that Eq.~(\ref{eincexp}) contains
also outgoing multipoles represented by spherical Hankel
function of the first kind ($h_l^{(1)}(\rho)$)~\cite{jackson99},
with the same amplitude of the incoming ones, i.e. no sources at the 
origin~\cite{sheppard97}.
That implies $h_l^{(1)}(\rho)+h_l^{(2)}(\rho)=2j_l(\rho)$,
where $j_l(\rho)$ is the spherical Bessel function of the first 
kind~\cite{jackson99}. We use the notation
$\mathbf{N}_{\mathrm{e}1l}^{(1)}$ and $\mathbf{M}_{\mathrm{o}1l}^{(1)}$ for 
multipoles with $j_l(\rho)$ and include the factor 2 into $A_l$ to 
eliminate it from Eq.~(\ref{eincexp}).

When a spherical NP of radius $a$ is placed at the focus,
we have to solve for a boundary value problem involving
the incident field $\mathbf{E}_\mathrm{i}$,
the scattered field $\mathbf{E}_\mathrm{s}$ and the internal
field $\mathbf{E}_\mathrm{int}$. The sum of incident and scattered
fields builds up the total field
$\mathbf{E}_\mathrm{tot}=\mathbf{E}_\mathrm{i}+\mathbf{E}_\mathrm{s}$
in the background region\cite{bohren83}. 
The symmetry of the problem implies that the NP will
not excite multipoles except those present in the incident field.
Therefore, we can formally write the solution as a linear
superposition of multipoles
\begin{eqnarray}
\label{einc}
\mathbf{E}_\mathrm{i} & = & \sum_l A_l \left(
\mathbf{N}_{\mathrm{e}1l}^{(1)}-i
\mathbf{M}_{\mathrm{o}1l}^{(1)}\right), \\
\label{esca}
\mathbf{E}_\mathrm{s} & = & \sum_l A_l \left(-a_l
\mathbf{N}_{\mathrm{e}1l}^{(3)}+ib_l
\mathbf{M}_{\mathrm{o}1l}^{(3)}\right),\\
\label{eint}
\mathbf{E}_\mathrm{int} & = & \sum_l A_l \left(d_l
\mathbf{N}_{\mathrm{e}1l}^{(1)}-
ic_l\mathbf{M}_{\mathrm{o}1l}^{(1)}\right),
\end{eqnarray}
where the notation $\mathbf{N}_{\mathrm{e}1l}^{(3)}$ and 
$\mathbf{M}_{\mathrm{o}1l}^{(3)}$ means $h_l^{(1)}(\rho)$
in the multipoles since the scattered field must be an outgoing wave.
The magnetic field $\mathbf{H}$ is simply obtained from Maxwell's 
equations and from the properties of spherical vector 
harmonics~\cite{bohren83}.

Fulfilling the boundary conditions at the NP surface $r=a$ 
leads to an expression for the expansion coefficients that reads
\begin{eqnarray}
\label{miea}
a_l &=& \frac{m\psi_l(mx)\psi_l'(x)-\psi_l(x)\psi_l'(mx)}
{m\psi_l(mx)\chi_l'(x)-\chi_l(x)\psi_l'(mx)}, \\
\label{mieb}
b_l &=& \frac{m\psi_l(x)\psi_l'(mx)-\psi_l(mx)\psi_l'(x)}
{m\chi_l(x)\psi_l'(mx)-\psi_l(mx)\chi_l'(x)},
\end{eqnarray}
where $\psi_l(\rho)=\rho j_l(\rho)$ and  $\chi_l(\rho)=\rho 
h_l^{(1)}(\rho)$ are Riccati-Bessel functions, $x=ka$ is the size 
parameter and $m=k_1/k$ is the relative refractive 
index~\cite{bohren83}, if $k_1$ is the wavevector inside the NP. 
Similar expressions hold for the coefficients of the internal field
\begin{eqnarray}
\label{miec}
c_l &=& \frac{m\chi_l(x)\psi_l'(x)-m\psi_l(x)\chi_l'(x)}
{m\chi_l(x)\psi_l'(mx)-\psi_l(mx)\chi_l'(x)}, \\
\label{mied}
d_l &=& \frac{m\psi_l(x)\chi_l'(x)-m\chi_l(x)\psi_l'(x)}
{m\psi_l(mx)\chi_l'(x)-\chi_l(x)\psi_l'(mx)}.
\end{eqnarray}
Notice that the coefficients of Eqs.~(\ref{miea}-\ref{mied})
have exactly the same expression as the Mie coefficients for an incident 
PW~\cite{bohren83}.
Once they are known, we simply use Eqs.~(\ref{einc}-\ref{eint}) to 
compute the electromagnetic field inside and outside the NP 
analytically.

\subsection{Far-field Cross Sections}

The interaction of light with material particles is usually investigated 
by considering far-field cross sections.
The scattering and absorption cross sections, $C_\mathrm{s}$ and 
$C_\mathrm{a}$, represent how strongly a 
particle scatters or absorbs light, respectively, for a given incident 
intensity. The total amount of power that is removed from the incident 
beam via scattering and absorption is proportional to the extinction 
cross section $C_\mathrm{e}=C_\mathrm{s}+C_\mathrm{a}$~\cite{bohren83}.

Since the radial component of every multipole vanishes in the far field,
the scattered power $W_\mathrm{s}$ reads
\begin{equation}
\label{wsca}
W_\mathrm{s}=\lim_{r\to\infty}\frac{1}{2}\int_0^{2\pi}\int_0^\pi
\mathrm{Re}\left\{E_{\mathrm{s},\theta}H_{\mathrm{s},\phi}^*-
E_{\mathrm{s},\phi}H_{\mathrm{s},\theta}^*\right\}
r^2\sin\theta\mathrm{d}\theta\mathrm{d}\phi,
\end{equation}
where 
$\mathbf{E}_\mathrm{s}=(E_{\mathrm{s},r},E_{\mathrm{s},\theta},
E_{\mathrm{s},\phi})$ and $\mathbf{H}_\mathrm{s}=(H_{\mathrm{s},r},
H_{\mathrm{s},\theta},H_{\mathrm{s},\phi})$ represent the 
scattered field.
After performing the integration over the angles and exploiting the 
orthogonality of the angle-dependent functions 
$\tau_l=\mathrm{d}P_l^1(\cos\theta)/\mathrm{d}\theta$ and
$\pi_l=P_l^1(\cos\theta)/\sin\theta$~\cite{bohren83}, Eq.~(\ref{wsca}) 
simplifies to
\begin{equation}
\label{wscas}
W_\mathrm{s}=\frac{\pi}{2Zk^2}\sum_l|A_l|^2
\frac{2l^2(l+1)^2}{2l+1}\left(|a_l|^2+|b_l|^2\right),
\end{equation}
where $Z$ is the background medium impedance. Likewise, the extinguished 
power $W_\mathrm{e}$ is defined as
\begin{equation}
\label{wext}
W_\mathrm{e}=\lim_{r\to\infty}\frac{1}{2}\int_0^{2\pi}\int_0^\pi
\mathrm{Re}\left\{E_{\mathrm{i},\phi}H_{\mathrm{s},\theta}^*-
E_{\mathrm{i},\theta}H_{\mathrm{s},\phi}^*-E_{\mathrm{s},\theta}
H_{\mathrm{i},\phi}^*+E_{\mathrm{s},\phi}
H_{\mathrm{i},\theta}^*\right\}
r^2\sin\theta\mathrm{d}\theta\mathrm{d}\phi,
\end{equation}
where 
$\mathbf{E}_\mathrm{i}=(E_{\mathrm{i},r},E_{\mathrm{i},\theta},
E_{\mathrm{i},\phi})$ and $\mathbf{H}_\mathrm{i}=(H_{\mathrm{i},r},
H_{\mathrm{i},\theta},H_{\mathrm{i},\phi})$ refer to the 
incident field. After integration Eq.~(\ref{wext}) reduces to
\begin{equation}
\label{wexts}
W_\mathrm{e}=\frac{\pi}{2Zk^2}\sum_l|A_l|^2\frac{2l^2(l+1)^2}{2l+1}
\mathrm{Re}\left\{a_l+b_l\right\}.
\end{equation}
Notice that Eqs.~(\ref{wscas}) and (\ref{wexts}) have the same 
dependence on Mie coefficients as for a PW~\cite{bohren83}.
Since the intensity of a PW is homogeneous, the cross sections
are easily obtained by dividing the scattered or extinguished powers
by the illumination intensity. However, for a high-NA optical system the 
intensity at the focus is inhomogeneous, and introducing a cross section 
is not straightforward.
Nevertheless, in order to be able to compare our results with that of a
PW~\cite{bohren83}, we define the cross section by dividing Eq.~(\ref{wscas})
and (\ref{wexts}) by the incident intensity $I_\mathrm{i}$,
\begin{equation}
\label{avint}
I_\mathrm{i}=\frac{1}{2\pi a^2} \int_A \mathrm{Re}
\left\{ \mathbf{E}_\mathrm{i} 
\times \mathbf{H}_\mathrm{i}^\ast \right\}_{-z} \mathrm{d}s.
\end{equation}
Here we compute the Poynting vector along $-z$ and take its average over 
the NP $xy$ circular section of area $A=\pi a^2$ at the focal plane.

\subsection{Average Intensity Enhancement}

Far-field quantities like the cross sections do not provide information
on the near field. To further compare the optical response of
a NP illuminated by a high-NA beam with a PW,
we would like to consider a parameter similar to the cross section
as an estimate of the complex structure of the near field.
Messinger et al.~\cite{messinger81} have defined
the near-field quantity $Q_\mathrm{NF}$ by computing the scattering 
cross section at $r=a$, including the radial field component, and 
dividing it by the NP area.
They also considered a similar quantity, called $Q_\mathrm{R}$, by 
taking only the radial component of the scattered field.
They found that $Q_\mathrm{R}$ is very close to $Q_\mathrm{NF}$
because in the near-field the radial component is quite large.
However, to study the near-field more precisely, one has to take
into account both the incident and the scattered fields. 
Thus we evaluate the total field intensity 
$|\mathbf{E}_\mathrm{tot}|^2$, average it over the NP surface and 
then normalize it by the intensity at the focus to obtain an 
enhancement factor,
\begin{equation}
\label{Ken}
K=\frac{1}{4\pi \left|\mathbf{E}_i(0,0,0)\right|^2}
\int_0^{2\pi}\int_0^\pi
\left| \mathbf{E}_\mathrm{tot}(a,\theta,\phi) \right|^2 
\sin\theta\mathrm{d}\theta\mathrm{d}\phi .
\end{equation}
The same average intensity enhancement can be computed for
PW illumination to compare the two situations.
It turns out to be more instructive if one looks at the
radial and tangential components of the total field separately.
Therefore, we split $K$ into a radial $K_\mathrm{r}$ and tangential
$K_\mathrm{t}$ average intensity enhancement.
By taking advantage of the orthogonality properties of multipoles,
these quantities have the analytical expression
\begin{eqnarray}
\label{Kren}
&& K_\mathrm{r} = \frac{9}{16 (ka)^4} \sum_l
\frac{\left|A_l\right|^2}{\left|A_1\right|^2}\frac{2l^3(l+1)^3}{2l+1}
\left[\left|a_l\right|^2 \left|\chi_l\right|^2 +\left|\psi_l\right|^2
-2\mathrm{Re}\left\{a_l \chi_l\right\} \psi_l\right], \\ 
\label{Kten}
&& \begin{split}K_\mathrm{t} = 
\frac{9}{16 (ka)^2} \sum_l 
\frac{\left|A_l\right|^2}{\left| A_1 \right|^2}
\frac{2l^2 (l+1)^2}{2l+1}
[\left| a_l\right|^2 \left|\chi_l' \right|^2 +
\left| b_l\right|^2 \left|\chi_l \right|^2 +\left|\psi_l\right|^2 +
\left| \psi_l' \right|^2 - \\
\left. 2\mathrm{Re}\left\{ a_l \chi_l' \right\} 
\psi_l' -2\mathrm{Re}\left\{ b_l \chi_l\right\}
\psi_l \right], \end{split} \\ \nonumber
\end{eqnarray}
where $\psi_l(\rho)$ and $\chi_l(\rho)$ are calculated at the
surface of the NP, that is $\rho=ka$. The coefficient $A_1$ at the 
denominator derives from the normalization
$|\mathbf{E}_\mathrm{i}(0,0,0)|^2=4|A_1|^2/9$. It can be shown that the 
same quantities for a NP illuminated by a PW 
have exactly the same form, except that $A_l$ is replaced by $E_l$, 
where $E_l=E_0i^l(2l+1)/(l(l+1))$ is the multipole-expansion coefficient
for a PW\cite{bohren83}.

When the NP is very small compared to the wavelength, only the
first multipole contributes to the intensity enhancement both
for a high-NA beam and a PW. By using the asymptotic expansions
of Mie coefficients and Riccati-Bessel functions for 
$\rho\to 0$~\cite{bohren83}, we obtain the following simple 
expressions,
\begin{eqnarray}
\label{Krenasym}
K_\mathrm{r} & = & \frac{1}{3}\left[ 4 \left|\alpha\right|^2+4 
\mathrm{Re}\left\{\alpha\right\} +1\right], \\
\label{Ktenasym}
K_\mathrm{t} & = & \frac{2}{3}\left[\left|\alpha\right|^2 -2 
\mathrm{Re}\left\{\alpha\right\} +1\right],
\end{eqnarray}
where $\alpha$ is a quantity proportional to the NP polarizability
per unit volume and is given by
\begin{equation}
\alpha=\frac{m^2-1}{m^2+2}.
\end{equation}
Equations~(\ref{Krenasym}-\ref{Ktenasym}) contain contributions from the 
scattered field, the incident field and their interference. Similar formulae
have been proposed to explain the electromagnetic contribution to 
surface-enhanced Raman scattering in the limit of very 
small NPs~\cite{moskovits85}.

\section{RESULTS AND DISCUSSION}
\label{results}

To understand the interaction of a high-NA beam with metal NPs,
we choose gold and silver NPs with a diameter of 100~nm embedded in
low-index lossless dielectric media, such as glass 
($n_\mathrm{b}=1.52$) and air ($n_\mathrm{b}=1$). 
The NP size is assumed to be smaller compared to the focal spot, but
sufficiently large to show deviations from PW illumination. Moreover,
while 100~nm gold NPs exhibit mainly a dipolar SPR, 100~nm silver
NPs can support higher-order SPRs. This allows us
to investigate the role of SPRs  in determining the 
optical response as a function of focusing.
The analytical formulae of Sec.~\ref{theory} have been computed
using Mathematica~\cite{mathematica}. Since the size parameter for
100~nm NPs illuminated in the wavelength range from 300~nm to
900~nm is close to one, we include up to 15 multipoles in the
field expansions. Including more multipoles did not bring a noticeable
change in our results. For modeling gold and silver NPs we use
the optical constants from Ref.~\onlinecite{johnson72}.

As it can be seen in Fig.~\ref{multipole}, if the beam gets focused 
more tightly, the dependence on higher-order multipoles becomes weaker 
compared to the case of PW illumination. Since the same multipole 
coefficients are present also in the scattered field, we expect that 
the excitation of higher-order SPRs will be strongly suppressed by
high-NA beams.

\subsection{Far-field Cross Sections}

\begin{figure}[h]
\begin{center}
\includegraphics[width =8.3cm]{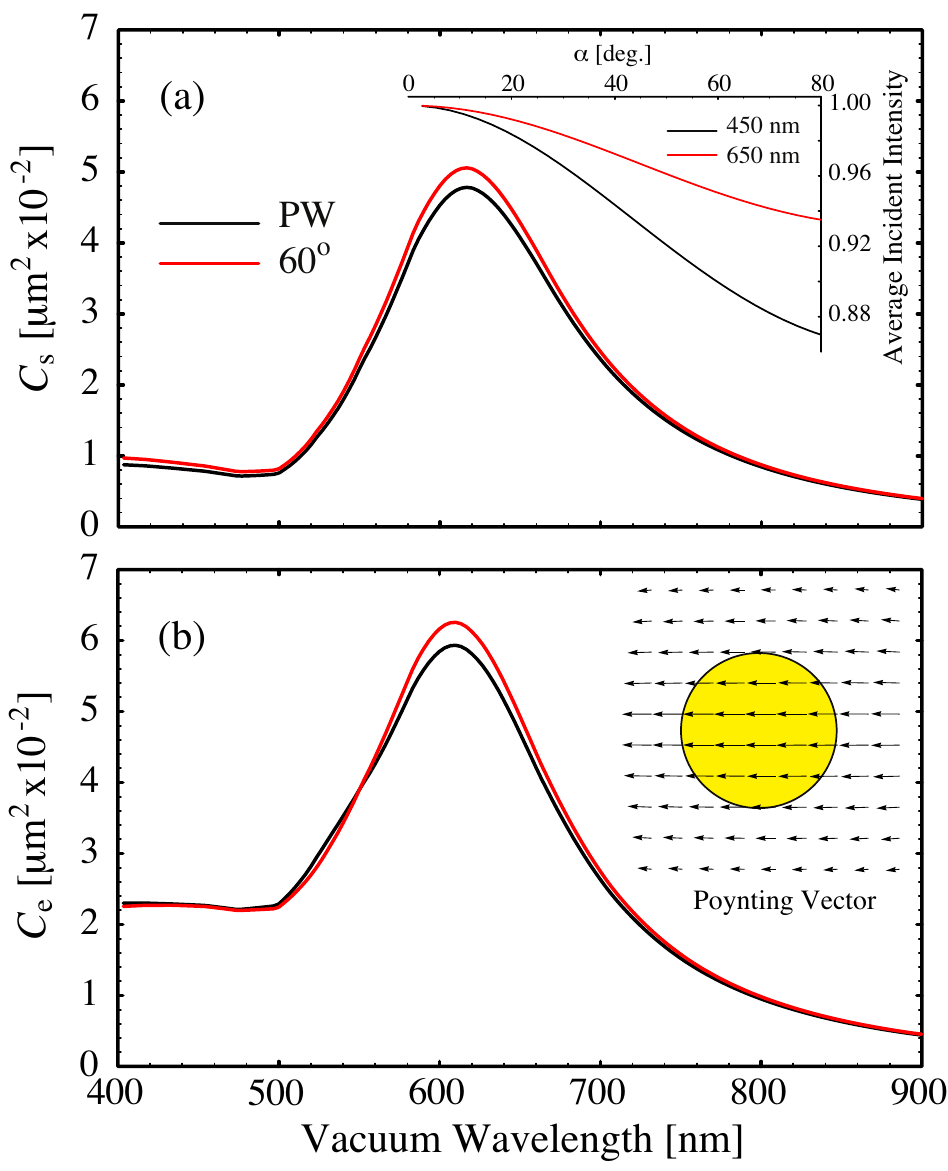}
\caption{\label{scsgold}
Scattering (a) and extinction (b) cross sections of a 
100~nm gold NP in glass illuminated by a PW and a high-NA
beam ($\alpha=60^\mathrm{o}$).
The inset to (a) shows the incident intensity averaged on the NP according
to Eq.~(\ref{avint}) and normalized with respect to the 
intensity at the origin as a function of the angular semi-aperture 
$\alpha$ for $\lambda=450$~nm and $\lambda = 650$~nm.
The intensiy at the origin is equal to that assumed for PW illumination.
The inset to (b) plots the Poynting vector for the high-NA beam 
($\alpha=60^\mathrm{o}$) at $\lambda=450$~nm with the 100~nm gold NP to 
scale.}
\end{center}
\end{figure}

Figure~\ref{scsgold} shows the scattering and extinction cross sections of 
a 100~nm gold NP in glass illuminated by a PW and a high-NA beam 
($\alpha=60^\mathrm{o}$).
Since the NP can support almost only a dipolar SPR, the
difference between the two illumination conditions is small.
The SPR peak is larger for a high-NA beam because the average incident 
intensity $I_\mathrm{i}$ is smaller than that of PW illumination, as shown 
in the inset to Fig.~\ref{scsgold}a.
If we had chosen the intensity at the origin for the normalization
of Eqs.~(\ref{wscas}) and (\ref{wexts}), we would have not seen an
increase in the dipolar peak.
For smaller gold NPs, the difference becomes negligible since the
average incident intensity is computed over a smaller area.
As shown in the inset, for a given NP and background medium the average
incident intensity depends on the wavelength and on $\alpha$.
The inset to Fig.~\ref{scsgold}b shows that the incident Poynting
vector is indeed completely along $-z$ across the NP, making the 
definition of average intensity according to Eq.~(\ref{avint}) a 
reasonable choice.

\begin{figure}[h]
\begin{center}
\includegraphics[width =8.3cm]{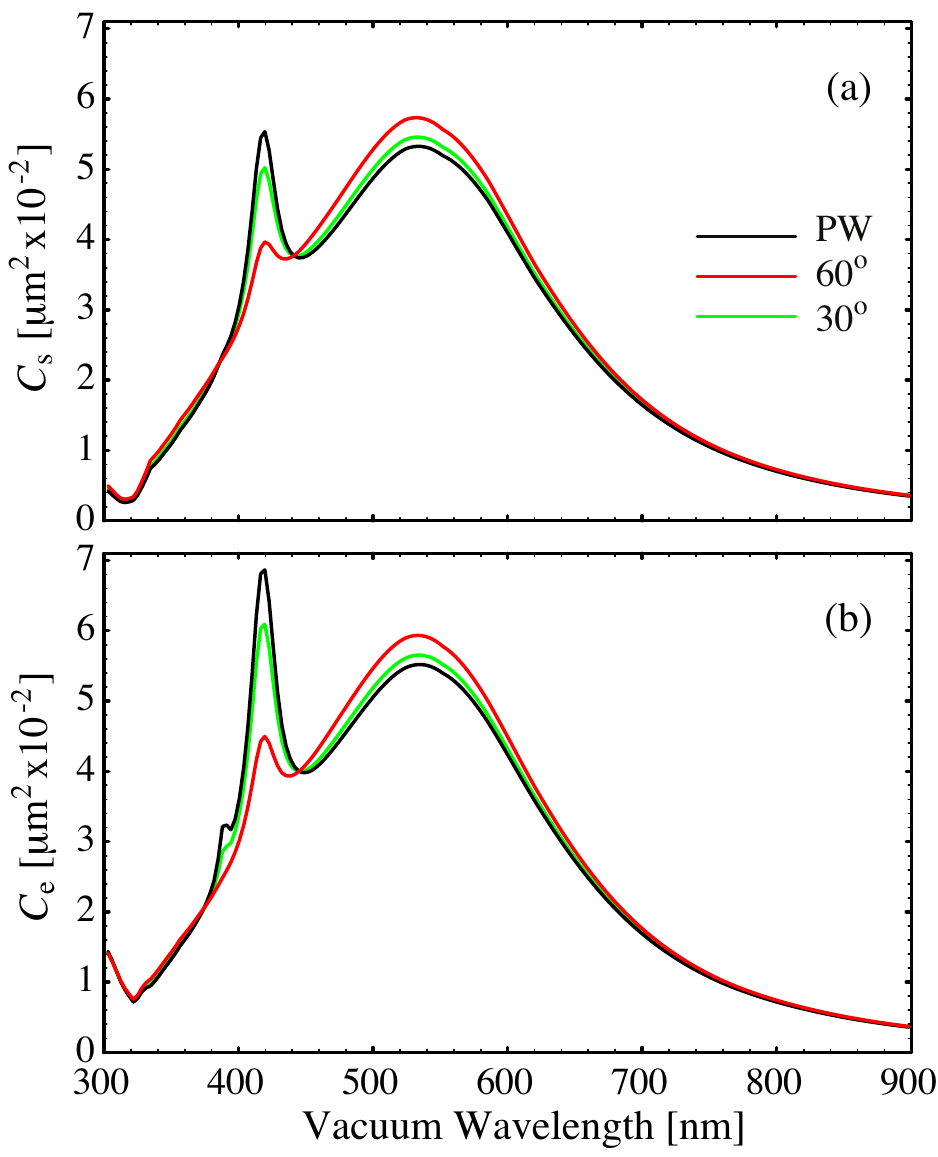}
\caption{\label{scssilver}
Scattering (a) and extinction (b) cross sections of a 
100~nm silver NP in glass illuminated by a PW and by a high-NA beam 
($\alpha=60^\mathrm{o}$ and $\alpha=30^\mathrm{o}$).}
\end{center}
\end{figure}

Figure~\ref{scssilver} shows the scattering and extinction cross 
sections of a 100~nm silver NP in glass
illuminated by a PW, a tightly ($\alpha=60^\mathrm{o}$) and a moderately 
($\alpha=30^\mathrm{o}$) focused beam.
Because the silver NP can support higher-order 
SPRs, the effect of focusing results in a strong suppression
of the quadrupole mode at $\lambda\simeq 420$~nm and of
the next higher-order mode at $\lambda\simeq 390$~nm. The latter
is visible only in the extinction cross section because
it mostly contributes to absorption rather than scattering.
As in the case of gold, the dipolar resonance at $\lambda\simeq 530$~nm
is enhanced due to the average incident intensity. The difference is
slightly larger than for gold because the resonance occurs at a shorter
wavelength, where the focal spot and the average incident
intensity are smaller (see inset to Fig.~\ref{scsgold}a).
Therefore, by adjusting the focus one can manipulate 
the excitation of SPRs in NPs and the optical response can be 
significantly changed even if the NP size is smaller
than the focal spot. For smaller particles the differences with respect to 
PW illumination vanish.

\subsection{Average Intensity Enhancement}

\begin{figure}[h]
\begin{center}
\includegraphics[width =8.3cm]{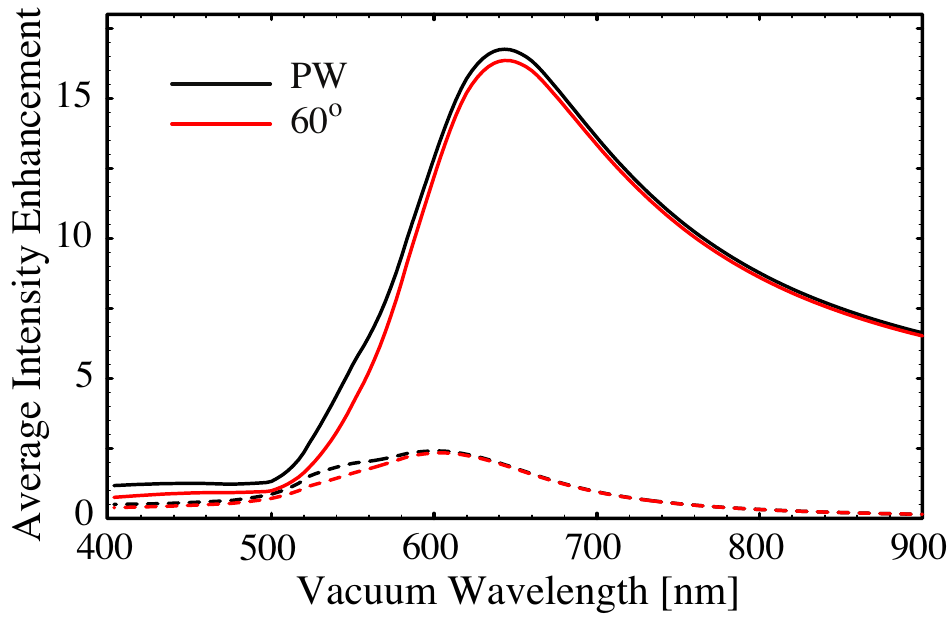}
\caption{\label{avengold}
Average radial (solid lines) and tangential (dashed lines) intensity 
enhancements evaluated at the surface of a 100~nm gold NP in glass 
illuminated by a PW and a high-NA beam ($\alpha=60^\mathrm{o}$).}
\end{center}
\end{figure}

We now focus our attention on the near field to see how the
field enhancement depends on the incident illumination. 
Figure~\ref{avengold} compares the average 
intensity enhancement of a 100~nm gold NP in glass.
Since the gold NP exhibits mainly a dipolar 
SPR, also in the near field the difference between 
focused beam and PW is small. In contrary to Fig.~\ref{scsgold}, the 
dipole peak is not enhanced by a focused beam because the normalization 
is with respect to the field at the origin.
In fact, a closer look at the data shows that the
average intensity enhancement is larger for a PW than
for a focused beam at the dipolar SPR. A small difference can be 
noticed also around $\lambda\simeq 550$~nm, where a weak quadrupole 
SPR exists. 
$K_\mathrm{r}$ is significantly larger than $K_\mathrm{t}$ as already 
found in Ref.~\onlinecite{messinger81}, where only the scattered 
field was considered.
It is interesting to notice that, as predicted by the terms in 
Eqs.~(\ref{Kren}) and (\ref{Kten}), the radial and the tangential
fields are maximal at different wavelengths.

\begin{figure}[h]
\begin{center}
\includegraphics [width =8.3cm]{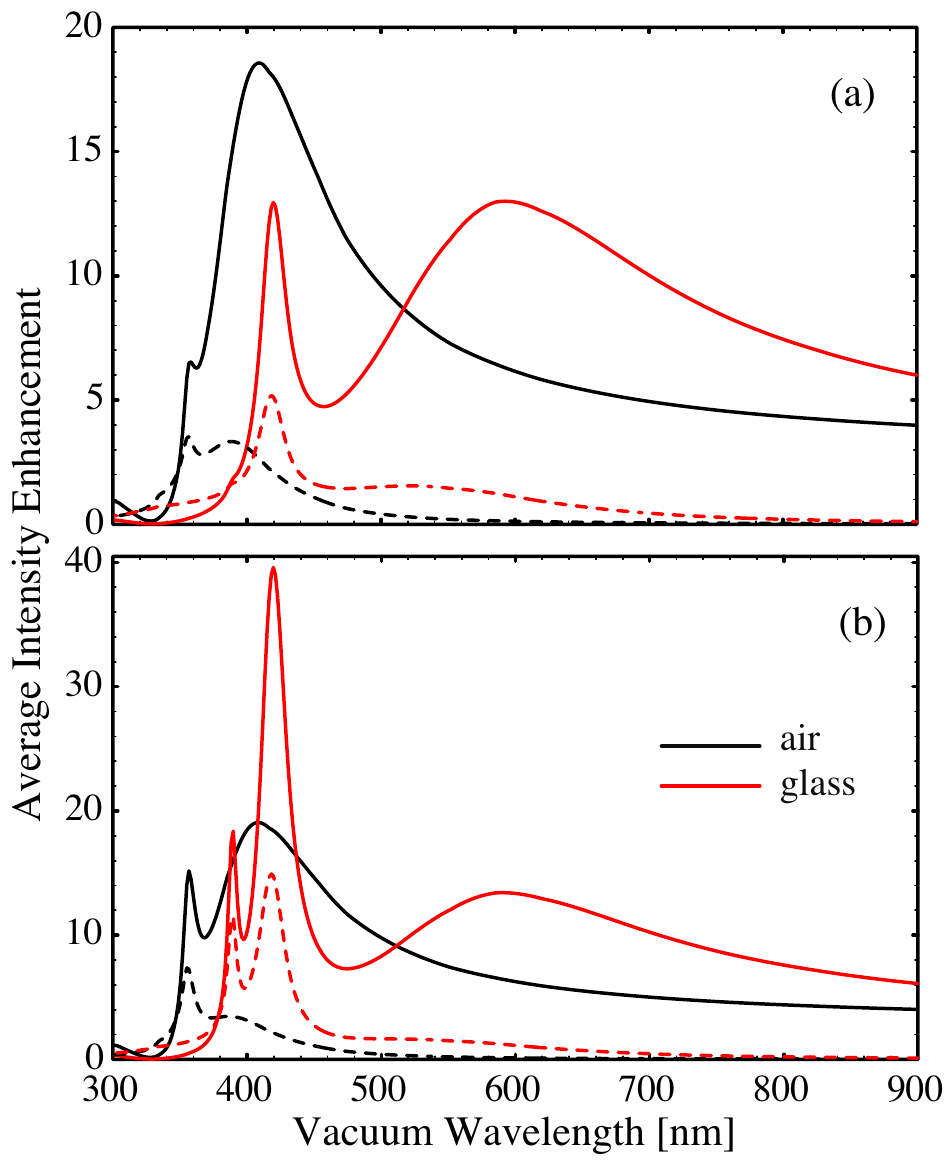}
\caption{\label{avensilver}
Average radial (solid lines) and tangential (dashed lines) intensity 
enhancements evaluated at the surface of a 100~nm silver NP in glass and air 
illuminated by a high-NA beam ($\alpha=60^\mathrm{o}$) (a) and a PW (b).}
\end{center}
\end{figure}

Figure~\ref{avensilver} shows that for a 100~nm silver NP the 
average intensity enhancement does depend on the strength of focusing.
Like the far-field cross sections, the higher-order SPRs are
weakly excited by a high-NA beam. Therefore, the enhancement
is much smaller compared to PW illumination in the spectral
region where the NP exhibits a quadrupole or a higher-order SPR.
On the other hand, the enhancement associated with the dipolar SPR
is left almost unchanged. Also for the silver NP, $K_\mathrm{r}$ is
larger than $K_\mathrm{t}$ even though for higher-order modes
the difference is less pronounced. It is also worth mentioning
that while for the dipolar SPR $K_\mathrm{r}$ and $K_\mathrm{t}$
are maximal at different wavelengths, they reach their maxima at
almost the same place for the higher-order SPRs.

The radial enhancement for the dipolar SPR is
red-shifted with respect to the dipolar peak in the far-field
cross sections for both gold and silver~\cite{messinger81}.
The fact that in going from air to glass the SPRs responsible 
for the near-field enhancement are red-shifted is
a well established result~\cite{bohren83}.
However, modes higher than the quadrupole are not excited by the 
high-NA beam (see Fig.~\ref{avensilver}).
Indeed, Fig.~\ref{multipole} confirms that multipoles of order $l$
larger than 2 are almost negligible in the beam expansion.
According to Eqs.~(\ref{Ktenasym}) and (\ref{Krenasym}), in the asymptotic 
limit of very small NPs there is no difference between PW and focused 
illumination because both contain the electric dipole multipole.

\subsection{Intensity Enhancement}

$K_\mathrm{r}$ and $K_\mathrm{t}$ provide us with a lot of information 
on the near field of metal NPs under different illumination conditions.
However, since they are average quantities,
they eliminate information on the spatial distribution of the 
enhancement. For this reason, we have to study the electric field 
pattern in the vicinity of the NP. We focus our attention on gold
at the peak wavelength of the dipolar SPR obtained from 
$K_\mathrm{r}$, since the higher-order 
SPRs in the silver NP would be weakly excited by the high-NA beam.
Moreover, we discuss only the radial component because the
tangential parts do not exhibit a large enhancement.
Again, it is worth mentioning that this quantity is normalized with 
respect to the intensity at the focus.

\begin{figure}[h]
\begin{center}
\includegraphics[width =8.3cm]{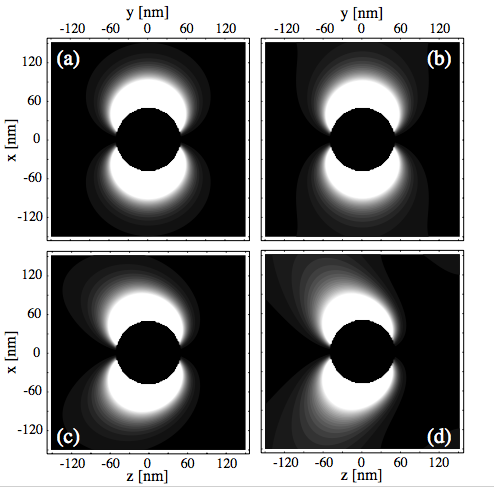}
\caption{\label{localen}
Radial intensity enhancement ($\log_{10}$ scale) for a 
100~nm gold NP in glass illuminated (along $-z$) by a high-NA beam 
($\alpha=60^\mathrm{o}$) at $\lambda=645$~nm ((a) and (c)) and by a PW 
at $\lambda=642$~nm ((b) and (d)). Cross cuts: $xy$ plane ((a) and (b)),
$xz$ plane ((c) and (d)). The field is not computed inside the NP.}
\end{center}
\end{figure}

Figure~\ref{localen} depicts the radial intensity enhancement
for a high-NA beam (panels (a) and (c)) and a PW 
(panels (b) and (d)) incident on a 100~nm gold NP in glass. As in the case 
of $K_\mathrm{r}$ we consider the radial component of the total field.
Notice that the wavelength chosen for the high-NA beam and the PW
are slightly different because the dipolar peaks in Fig.~\ref{avengold}
are not exactly at the same position. We have checked that
choosing the same wavelengths does not change the following results
since the SPRs are quite broad.
For the cross cuts corresponding to the focal plane (panels (a) and (b))
the field distribution looks very similar close to the NP surface, while
a deviation occurs away from it because the incident field starts to be
as strong as the scattered one.
The enhancement is maximal along the $x$ axis for both high-NA beam and 
PW. Also for the cross cuts in the $xz$ plane (panels (c) and (d)) the 
patterns are quite similar in proximity of the NP, but differences 
occur at larger distances. This can be explained by the 
different field profile of a high-NA beam and a PW.
As in the case of the far-field cross sections and the average
intensity enhancement, there is no remarkable difference
between a high-NA beam and a PW at the dipolar SPR.

\begin{figure}[h]
\begin{center}
\includegraphics[width=8.3cm]{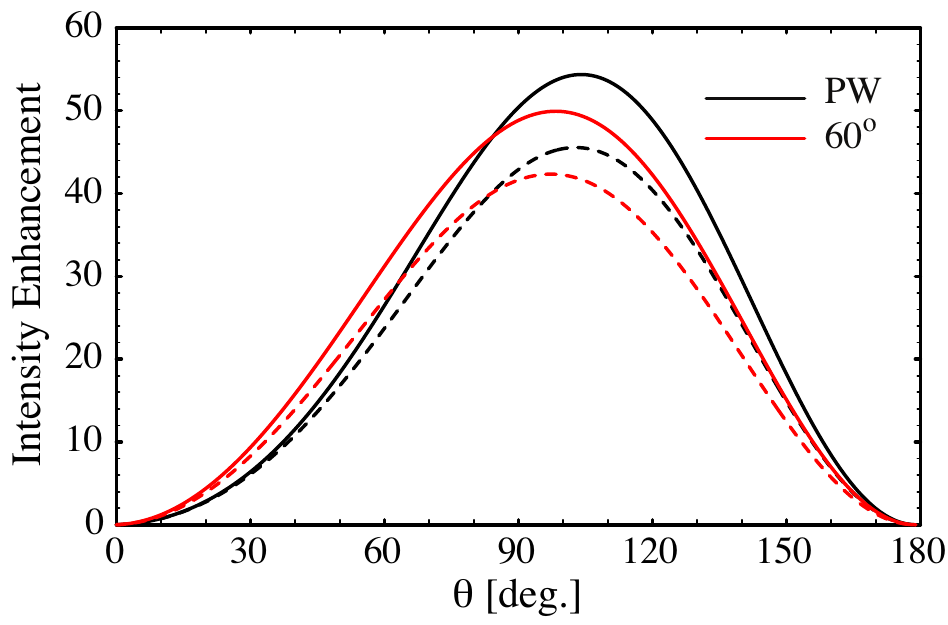}
\caption{\label{thetaen}
Radial intensity enhancement as a function 
of $\theta$ for $\phi=0^\mathrm{o}$ at the surface of a 100~nm gold NP 
in glass illuminated by a high-NA beam ($\alpha=60^\mathrm{o}$) at 
$\lambda=645$~nm and a PW at $\lambda=642$~nm. The contribution only
from the scattered field is represented by dashed lines.} 
\end{center}
\end{figure}

To have a closer look at the radial intensity enhancement
of Fig.~\ref{localen}, we plot it at the NP surface as a function
of $\theta$ and for $\phi=0^\mathrm{o}$, that is in the $xz$ plane.
Figure~\ref{thetaen} shows both the enhancement for the total intensity
(solid lines) and for only the scattered intensity (dashed lines).
First, it is seen that the constructive interference between incident 
and scattered field has a non-negligible contribution to the intensity 
enhancement. Second, the maximum for a high-NA beam and a PW are not
at the same location. For a high-NA beam the peak value is reached
at $\theta\simeq 100^\mathrm{o}$, while for a PW it is
reached at $\theta\simeq 105^\mathrm{o}$. Moreover, the overall
profile of the curves is slightly different. The enhancement for
a high-NA beam is more symmetric with respect to 
$\theta=90^\mathrm{o}$ and broader than that for a PW. 
On the other hand, the maximum enhancement is larger for a PW
than for a high-NA beam. We have checked that when the NP gets 
smaller these differences disappear and the maximum radial intensity 
enhancement occurs at $\theta=90^\mathrm{o}$ for both illuminations.

\section{CONCLUSIONS}

Using a multipole-expansion approach we have studied the interaction
of a high-NA beam with noble metal NPs placed at the focus
and have compared the results to the case of PW illumination.
The most important and general effect of focusing resides in the 
suppression of higher-order multipoles in the incident field expansion.
Consequently, even if the NP supports high-order SPRs, they
cannot be efficiently excited by a high-NA beam, as we have seen for the 
silver NP both in the far and near field. Therefore, care must be taken 
when interpreting the plasmon spectra measured under high-NA illumination.

In this work we have considered high-NA beams generated by
linearly-polarized light. It could be interesting to extend
these investigations to radially-polarized beams, which
achieve a tighter focus~\cite{quabis00,dorn03}.
These results might find application in areas like NP 
spectroscopy~\cite{krenn00,payne06}, optical data 
storage~\cite{ditlbacher00,sugiyama01} and optical 
forces~\cite{hallock05}.

\section*{Acknowledgments}
We thank V.~Jacobsen, and G.~Zumofen for fruitful 
discussions and P.~T\"or\"ok for a critical reading of the manuscript.
This work has been supported by ETH Zurich.

\end{document}